\documentclass[review]{elsarticle}

\usepackage{hyperref}

\journal{Journal of Medical Engeneering and Physics}

\begin{document}

\begin{frontmatter}

\title{1-D Convlutional Neural Networks for the Analysis of Pupil Size Variations in Scotopic Conditions}
\tnotetext[mytitlenote]{This work was partially  supported by Liquidweb and Roneuro as partners of the Neurosense Joint Laboratory.}



\author[mymainaddress]{Dario Zanca\corref{mycorrespondingauthor}}
\cortext[mycorrespondingauthor]{Corresponding author}
\ead{dario.zanca@unisi.it}

\author[mymainaddress]{Alessandra Rufa}
\ead{rufa@unisi.it}

\address[mymainaddress]{Neurosense Joint Laboratory, Department of Medicine, Surgery and Neuroscience, University of Siena, Siena, Italy}

\begin{abstract}
	It is well known that a systematic analysis of the pupil size variations, recorded by means of an eye-tracker, is a rich source of information about a subject’s arousal and cognitive state. Current methods for pupil analysis are limited to descriptive statistics, struggle in handling the wide inter-subjects variability and must be coupled with a long series of pre-processing signal operations. In this we present a data-driven approach in which 1-D Convolutional Neural Networks are applied directly to the raw pupil size data. To test its effectiveness, we apply our method in a binary classification task with two different groups of subjects: a group of elderly patients with Parkinson disease (PDs), a condition in which pupil abnormalities have been extensively reported, and a group of healthy adults subjects (HCs). Long-range registration (10 minutes) of the pupil size were collected in scotopic conditions (complete darkness, 0 lux). 1-D convolutional neural network models are trained for classification of short-range sequences (10 to 60 seconds of registration). The model provides prediction with high average accuracy on a hold out test set. Dataset and codes are released for reproducibility and benchmarking purposes.
\end{abstract}


\end{frontmatter}

\section{Introduction}\label{Introduction}
Pupils dilate and constrinct in response to at least three different kinds of stimuli~\cite{mathot2018pupillometry}: light, near fixations and increase in arousal and mental efforts.
Recent studies combining electrophysiology, optical imaging and neural networks modeling, indicated that the link between brain state activity and pupil size variations are related to the neuro-modulatory eﬀect of the noradrenergic and cholinergic systems~\cite{larsen2018neuromodulatory}.
Scotopic experimental conditions~\cite{crawford1949scotopic} (i.e. complete darkness, 0 lux) reveal that pupil size variations might be correlated to neural network oscillations~\cite{schwalm2017back}.
Several studies  in literature propose methods for   the analysis of the pupil size variations and for the characterization of this signal in isoluminant conditions or in the dark, with the goal of identifying patterns revealing changes in the cortical state activities. Under the assumption of linearity, common approaches usually involve short-time Fourier or wavelets transformations~\cite{nowak2014wavelet,nowak2008time,henson2010monitoring,reimer2014pupil} to give a signal representation.  This methods rely on an a priori choice of the basis function and often tacitly assume stationarity of the phenomenon, which hardly hold for many biometric psychological signals~\cite{poon1997decrease,regen2013association,morad2000pupillography}. Recently, some non-linear non-stationary approaches have been proposed that take advantage of projecting the pupil size signal  in a frequency domain~\cite{villalobos2016pupillometric}, in a latent space~\cite{mesin2013investigation,monaco2014evaluation}, or their  combination~\cite{piu2019cross}. In all cases, basis curves are function of time and an \textit{a posteriori} cross-correlation analysis allows to  investigate hypotheses on the dynamic interactions among the systems modulating pupil size variations. These approaches, based on signal's inspection, may have some limitations:  are difficult to extend to a more general modeling~\cite{zenon2017time}; do not deal with the wide variability between different subjects; require a large number of pre-processing operations (blink removal, normalization, filtering, among others). 

In this paper, we propose the learning of latent signal representations in a completely \textit{data-driven} approach, by minimizing a classification error function. Differently from other approaches~\cite{piu2019cross,mesin2013investigation,monaco2014evaluation}, we do not make use of \textit{ad hoc} normalizations to reduce inter-subject variability \textit{nor} for filtering noise, artifacts or outliers.  Data pre-processing is limited to the interpolation with cubic splines on missing data. Instead, a normalization rule is \textit{learned} in conjunction with the model, directly from data, by a properly defined normalization layer. 

We apply our method to the problem of classifying pupil raw signals belonging to two very different groups of subjects. The first group is composed by elderly patients diagnosed as suffering from Parkinson’s disease, the second group is composed of adult and healthy subjects. Our choice is motivated by the fact that Parkinson’s disease is a well known condition associated with changes in pupil size regulation~\cite{wang2016disruption}. 

\section{Method}\label{Method}

\subsection{Model definition}
We exploit 1-D convolutional neural network models~\cite{lecun1998gradient} (1D-CNN) for automatically extract effective features from the raw pupil size signal. Similar architectures are used in literature for similar problems involving the classification of other biometric 1-dimensional signals, like ECG~\cite{jambukia2015classification} or EEG~\cite{roy2019chrononet}. 

Feature extraction consists of layers realizing convolutional and pooling operations. Convolutional layers can be seen as the application of filters that enhance some features of the original signal while reducing noise. Finally a non-linear activation function is applied. The output of a convolutional layer can be written as $$x_j^l = \sigma(\sum_{i \in M_j} x_i^{l-1}*W_{i,j}^l + b_j^l),$$
where $M_j$ represents the receptive field of the current unit, $l$ is the layer index, $W_{i,j}^l$ are the parameters associated with the kernel applied and $\sigma$ is the non-linear activation function chosen for the layer. We indicate with $*$ the convolutional operator. 
At the top of the convolutional module, a number of fully connected layers is added,
which allows learning non-linear combinations of the convolutional features. The best number of convolutional and fully connected layers has been validated experimentally (with grid search). Finally a softmax layer generates a class prediction. 

Two main factors make it difficult to apply machine learning techniques directly to the raw pupil size data. The first is the scarcity of data, which takes a long time to be collected and the use of compatible hardware. The second is the wide variability of this type of data between different subjects. To overcome these problems we introduce two characteristics in the computational graph. First, we  introduce artificial noise by applying a dropout operation~\cite{srivastava2014dropout} to the input of the model. This is demonstrated to reduce the over-fitting by preventing complex co-adaptations on training data. Second, we explicitly learn a normalization rule in conjunction with the neural model. This is done by introducing a normalization layer~\cite{ioffe2015batch}. 
In the present literature, normalization of the pupil size signal is usually performed as pre-processing of the data, in order to reduce inter-subject variability~\cite{piu2019cross,mesin2013investigation,monaco2014evaluation}. We show that the learning of a normalization rule brings relevant improvements in the results.

\section{Experiments}\label{Experiments}

\subsection{Data collection}\label{Data collection}
The dataset contains the monocular pupil size recordings of 21 HCs (average age 36 $\pm$ 13) and 15 elder PDs (average age 69 $\pm$ 7), for a total of 36 participants. 
The pupil size recordings were acquired with an ASL 504 eye-tracker device (mean sampling frequency of 240 Hz). The distance of the camera from the participant's eye is 650 mm. The participant’s head is kept still by means of a chinrest for the entire experimental session. Data is collected in a  light-controlled setup in a scotopic condition (complete darkness, 0 lux measured)~\cite{crawford1949scotopic}. Participants are asked to look straight, minimize mental activity and relax. Participant pupil size is recorded for 11 minutes. The first minute of dark adaptation is discarded. Raw data is not filtered, nor normalized. 

In order to avoid the risk that performance can be traced back to noise patterns rather than true data, we also randomly superimpose noise patterns from one class on another.

Pupil size registrations are divided into sub-sequences, each corresponding to $n$ seconds of registration, with $n \in \{10, 15, 30, 60\}$. For the case of $10$ and $15$ seconds, we discard sequences that contain more than the $10\%$ of missing data. For longer sequences of $30$ and $60$ seconds, we increase this threshold to $15\%$ and $20\%$ respectively. Sub-sequences are then shuffled inside each set. For each of the conditions, we perform a 5-fold cross validation. The recordings are randomly divided into 5 folds as follows:  train (24 subjects), validation (6 subjects, balanced classes) and test sets (6 subjects, balanced classes). The folds are fixed in advance and the same folds are used for all trials. The folds configuration used in this paper is released together with the dataset for reproducibility purposes.

\subsection{Baselines}
Normalized graphs are calculated both for HCs and PDs. 
They are obtained by dividing pupil size subject-wise by the mean pupil size~\cite{piu2019cross}, for each individual registration. 

Normalized graphs are used to define and evaluate two baseline to be compared with the model's performance. For baseline \textit{B1}, we calculate the Euclidean distance of the normalized pupil size signal from the normalized graphs. The sample is assigned to the class corresponding to the nearest normative graph. For the second baseline \textit{B2} we compute the Kullback-Leibler divergence between two signals when viewed as distributions: it is a non-symmetric measure of the information lost when the normative graph is used as estimate of a given signal. Scores for the baselines are summarized in table~\ref{tab:baselines}.

        \begin{table}
        	\begin{center}
        		\begin{tabular}{cc}
        			\hline
            			\textbf{Baseline} & \textbf{Accuracy}\\ 
        			\hline
        			\hline
        			    $B1$ & $63.89\%$\\ 
        			\hline
        			    $B2$ & $61.11\%$\\ 
        			\hline
        		\end{tabular}
        		\caption{\textbf{Baselines $B1$ and $B2$.} Baseline \textit{B1} is based on the Euclidean distance of the normalized pupil size signal from the normalized graphs. The sample is assigned to the class corresponding with the nearest normative graph. For the second baseline \textit{B2} we compute the Kullback-Leibler divergence between two signals when viewed as distributions.}
        		 \label{tab:baselines}
        	\end{center}
        \end{table}

\subsection{Results}\label{Results}
We train the proposed 1D-CNNs models for a task of binary classification (i.e., HCs vs. PDs). We select a model architecture with a grid-search to identify the number of convolutional layers $n_c \in \{1,2,3\}$ and fully connected layers $n_{fc} \in \{1,2,3\}$. The best architecture for the task is defined in the table~\ref{tab:bestestimator}. As a reference, the proposed model is compared with a multi-layer perceptron network, see table~\ref{tab:mlp} for details. 

\begin{table}
        	\begin{center}
        	\begin{tabular}{cc}
        			\hline
            			 \textbf{Layer (type)} & \textbf{Units} \\ 
           			\hline
           			\hline
           			Batch norm. & Same as input\\
           			\hline
           			Dropout & $-$\\
           			\hline
           			Fully connected & $128$ $HU$\\
           			\hline
           			Fully connected & $64$ $HU$\\
           			\hline
           			Softmax & $2$\\
           			\hline
        		\end{tabular}
        		\caption{\textbf{MLP architecture.} As a reference, the proposed model is compared with a multi-layer perceptron network. It it composed by a batch normalization layer, and two fully connected layer. A softmax layer on the top. $HU$ = "hidden units".}
        		 \label{tab:mlp}
        	\end{center}
        \end{table}

\begin{table}
        	\begin{center}
        	\begin{tabular}{cc}
        			\hline
            			 \textbf{Layer (type)} & \textbf{Units} \\ 
           			\hline
           			\hline
           			Batch norm. & Same as input\\
           			\hline
           			Dropout & $-$\\
           			\hline
           			Convolutional & $16 F$,  $5 \times 1$\\
           			\hline
           			Pooling & $2 \times 1$\\
           			\hline
           			Convolutional & $32 F$,  $3 \times 1$\\
           			\hline
           			Flatten & $-$\\
           			\hline
           			Fully connected & $10$ $HU$\\
           			\hline
           			Softmax & $2$\\
           			\hline
        		\end{tabular}
        		\caption{\textbf{1D-CNN architecture.} The best estimator structure has been selected with a grid search. It it composed by a batch normalization layer, two convolutional layers and a fully connected layer. A softmax layer on the top. $F$ = "filters", $HU$ = "hidden units".}
        		 \label{tab:bestestimator}
        	\end{center}
        \end{table}

Table~\ref{tab:scoresmixed} shows mean accuracy and standard deviation in a 5-fold cross validation. In order to verify the effectiveness of the normalization layer, we compare the performance with an identical 1D-CNN in which the normalization layer \textit{BN} is substituted by a pre-processing operation \textit{PP} of data normalization. In particular, sequences are re-scaled by dividing by the subject's mean pupil size. 
Because of the high inter-subject variability (and scarceness of samples), the network is not able to perform in a comparable way without a parameterized and learnable normalization layer. 
The  model's best performance are obtained for 15-seconds sequences. This may find an explanation in the fact that longer sequences are more affected by the noise given by the missing data. Also, as the input dimensionality grows, so the feature space dimensionality does, while the size of the dataset decreases. Due to the very spatial nature of the problem, a classical MLP is not able to learn the classification task.

 \begin{table}
        	\begin{center}
        	\begin{tabular}{cccc}
        			\hline
            			 & \textbf{Seq. len.} & \textbf{Norm.} & \textbf{Accuracy} \\ 
            			\textbf{Model} & \textbf{(sec.)} & \textbf{(PP/BN) }& \textbf{($\%$)} \\ 
        			\hline
        			\hline
        			    MLP & 10 &PP& $50.20 (\pm 1.66)$\\ 
        			    MLP & 10 &BN& $56.00 (\pm 0.03)$\\ 
        			    1D-CNN & 10 &PP& $54.20 (\pm 1.56)$\\ 
        			    1D-CNN & 10 &BN& $\textbf{77.19} (\pm \textbf{3.33})$\\ 
        			\hline
        			\hline
        			    MLP & 15 &PP& $51.15 (\pm 0.9)$\\ 
        			    MLP & 15 &BN& $57.23 (\pm 0.02)$\\ 
        			    1D-CNN & 15 &PP& $74.65 (\pm 1.08)$\\ 
        			    1D-CNN & 15 &BN& $\textbf{81.26} (\pm \textbf{1.79})$\\ 
        			\hline
        			\hline
        			    MLP & 30 &PP& $50.87 (\pm 0.81)$\\ 
        			    MLP & 30 &BN& $54.11 (\pm 0.09)$\\ 
        			    1D-CNN & 30 &PP& $56.67 (\pm 1.77)$\\ 
        			    1D-CNN & 30 &BN& $\textbf{79.27} (\pm \textbf{3.99})$\\ 
        			\hline
        			\hline
        			    MLP & 60 &PP& $50.95 (\pm 1.15)$\\ 
        			    MLP & 60 &BN& $55.23 (\pm 0.55)$\\ 
        			    1D-CNN & 60 &PP& $59.99 (\pm 3.99)$\\ 
        			    1D-CNN & 60 &BN& $\textbf{72.20} (\pm \textbf{3.37})$\\ 
        			\hline
        		\end{tabular}
        		\caption{\textbf{Accuracy on pupil signal classification (mixed missing data).} Model performance are summarized in this table for the case of the dataset with mixed missing data. The last column report the average accuracy score on the test set for a 5-fold cross validation. Standard deviation is between brackets. Again, two types of normalization have been evaluated: PP (pre-processing) scales data dividing the the subject's mean pupil size, BP (batch normalization) learns normalization in conjunction with the network as a normalization layer. In bold, the best predictor's score.}
        		 \label{tab:scoresmixed}
        	\end{center}
        \end{table}


\section{Conclusion}\label{Conclusion}
In this paper we provide a twofold contribution. On the one hand, a dataset of raw pupil data of two very different groups of subjects (adult HCs and elder PDs) collected in wakefulness scotopic conditions in order to enhance the contribution of cortical state activations to the pupil size variations and minimizing the effects of light and near fixations. On the other hand, a tool for the automatic learning of pupil size features using 1D-CNNs is learned from the data.  



Future works may include learning long-term dependencies through the use of recurring architectures. To this end, larger amount of data needs to be collected. Since the data of healthy subjects are more easily accessible, this would lead to very unbalanced datasets. However, techniques can be imported from the literature of novelty detection or few shot learning to extend the methodology proposed in this manuscript.


\section*{Acknowledgement}\label{References}
We thank Silvio Sabatini and Andrea Canessa (University of Genova) for insightful discussions and important suggestions. The work was supported by RoNeuro and Liquidweb srl who participate in the Neurosense joint  lab of the University of Siena. 

Conflicts of Interest: None. Funding: None.  Ethical Approval: The study was approved by the local Ethical Committee Comitato Etico Locale Azienda Ospedaliera Universitaria Senese, EVAlab protocol CEL no. 48/2010.

\bibliography{mybibfile}

\end{document}